\documentclass[aps,prb,twocolumn,showpacs,superscriptaddress,groupedaddress]{revtex4}  
\usepackage{graphicx}  
\usepackage{dcolumn}   
\usepackage{bm}        
\usepackage{amssymb,cases}   

\newcommand{\hbq}{\hat{\mathbf q}}

\newcommand{\hbo}{\hat{\boldsymbol\omega}}
\newcommand{\hbz}{\hat{\mathbf z}}

\newcommand{\lk}{\left (}
\newcommand{\rk}{\right )}

\newcommand{\qa}{\begin{quote}}
\newcommand{\qe}{\end{quote}}

\newcommand{\ma}{\begin{pmatrix}}
\newcommand{\me}{\end{pmatrix}}

\newcommand{\ba}{{\begin{align}}}
\newcommand{\ea}{{\end{align}}}
\newcommand{\ee}{\end{eqnarray}}
\newcommand{\eql[1]}{\marginpar{\vspace{.5cm}\tiny{\{#1}\}}\begin{eqnarray}\label{#1}}
\newcommand{\eqa}{\begin{align}}
\newcommand{\eqan}{\begin{eqnarray*}}
\newcommand{\eqe}{\end{align} }
\newcommand{\sqa}{\begin{subequations}
\begin{empheq}{align}}
\newcommand{\sqe}{\end{empheq}
\end{subequations}}
\newcommand{\nn}{\nonumber}

\renewcommand{\c}{\mathcal }

\renewcommand{\b }{\bs }

\newcommand{\bs}{\boldsymbol}

\newcommand{\hbe}{ {\hat{\bs e}}}

\newcommand{\bnabla}{\bs\nabla}

\newcommand{\br}{{\bs r}}

\begin{document}

\widetext
\leftline{JETP, November 2018}
\newcommand{\hpq}{\hat{\mathbf q}}

\title{Topological defects in helical magnets}
\author{T. Nattermann} 
\affiliation{Institute for Theoretical Physics, University of Cologne, 50937 Cologne, Germany}                             
 \author{V.L. Pokrovsky}                            
\affiliation{Department of Physics, Texas A\&M University,
College Station, TX 77843-4242} 
\affiliation{Landau Institute for Theoretical Physics,
Chernogolovka, Moscow District 142432, Russia}                             
\date{\today}

\begin{abstract}
Helical magnets which violated space inversion symmetry have rather peculiar topological defects. In isotropic helical magnets with exchange and Dzyaloshinskii-Moriya interactions  there are only three types of linear defects: $\pm\pi$ and $2\pi$-disclinations. Weak crystal anysotropy suppresses linear defects on large scale. Instead planar defects appear: domain walls that separate domains with different preferential directions of helical wave vectors. The appearance of such domain walls in the bulk helical magnets and some of their properties were predicted in the work \cite{Li 2012}. In a recent work by an international team of experimenters and theorists \cite{Schoenherr 2018} the existence of new types of domain walls on crystal faces of helical magnet FeGe was discovered. They have many features predicted by theory \cite{Li 2012}, but display also unexpected properties, one of them is the possibility of arbitrary angle between helical wave vectors. Depending on this angle the domain walls observed in \cite{Schoenherr 2018} can be divided in two classes: smooth and zig-zag. This article contains a mini-review of the existing theory and experiment. It also contains new results that explain why in a system with continuos orientation of helical wave vectors domain walls are possible. We discuss why and at what conditions smooth and zig-zag domain walls appear, analyze spin textures associated with helical domain walls and find the dependence of their width on angle between helical wave vectors. 
\end{abstract}

\maketitle

\section{Introduction}

We are happy to congratulate Lev Pitaevskii on occasion of his Jubilee. 
We are amazed by the depth and brilliance of his scientific achievements
that include such pearls as the {\it Gross-Pitaevskii equation} and the {\it Theory
    of electromagnetic fluctuations in dispersive media}. But not less
surprising is his universalism, so rare in our time, and his irreproachable 
scientific integrity. These qualities together made him a unique person,
able to continue the magnificient Landau-Lifshitz compendium of modern
theoretical physics. The fact that it is still not simply alive after
the death of both inital authors, but is an indispensable belonging 
of any private and institutional physical library, is the result of his tireless work.
One of us (VP) has the privilege and pleasure to be in friendship with
him starting from the end of 1950-th. He learned a lot from scientific
discussions with Lev but not only that. Lev's natural kindness and
sincerity is an important part of life for all his friends. We wish
him good health and high spirit. 

In this article, written in his honour, we 
review theory and experiment 
on topological defects in helical magnets. Topological defects are
among the most facscinating objects of condensed matter physics and
quantum field theory \cite{Toulouse 1976,Volovik 1977,Chaikin 1995,Volovik 2003,Polyakov 1987}.
They almost unavoidably appear in ordered phases, either as a result
of initial conditions like cosmic strings {\cite{Polyakov 1987,Volovik 2003},
    or as equilibrium configurations like vortex lattices in type-II superconductors
    \cite{Abrikosov 1957} and superfluids \cite{Onsager 1949,Feynman 1955}.
    Defects counteract the emergent rigidity of the condensate and hence
    are fundamental both from the point of view of basic science as well
    as of practical applications\cite{Haasen 1971,Anderson 1984}
    
    Topological defects are different in systems with different symmetries
    and types of its violation in the ordered states. Some of
    them are well known, for example quantized vortices in quantum liquids
    \cite{Onsager 1949,Abrikosov 1957} or domain walls in magnets. Skyrmions
    \cite{Skyrme 1962,Belavin 1975, Abanov 1998} are more sophisticated but obtained
    a broad advertisement recently due to experiments in 2-dimensional
    magnets and nuclear matter\cite{Bogdanov 1994,Muhlbauer 2009,Yu 2010, Kiselev 2011, Fert 2017}.
    Defects in superfluid helium 3 with its complex ordering are more
    exotic. Their description and literature can be found in the book by Volovik \cite{Volovik 1992}. 
    
    Topological defects in 
    helical magnets are interesting not only
    as a subject of pure science, but also because they interact with
    electric current and thus may serve for transformation of magnetic
    signals into electric ones and back, that is a basic element of sensing,
    transfer and storing of information and energy. 
    
    In the year 2012 Fuxiang
    Li and the authors published an article on domain walls in helical
    magnets, in which we predicted rather unusual properties of these defects \cite{Li 2012}.
    Most of these predictions were confirmed in a fundamental experiments
    performed by the international team of experimenters and theorists 
    in 2018 \cite{Schoenherr 2018},
    but not all and some unexpected features appeared. In the following
    text we analyze these experiments and explain some of discrepancies
    between our theory and the experimental observations. Explanations of other facts require a further development of theory.
    The main reason of discrepancies is that Magnetic Force Microscope
    (MFM), used in these experiments, gives information on magnetic textures
    on crystal faces of samples, whereas our theory had in mind the bulk. 
    
    The content of the following article is as follows. In the second
    section we present a general analysis of the topological defects in isotropic
    helical magnets. 
    In the third section we consider in more details domain walls
    in the bulk helical magnets. The fourth section is dedicated to helical domain walls
    at crystal faces. It contains a brief description of the experimental observations
    made in the work \cite{Schoenherr 2018} and modifications of the domain walls theory necessary 
    in this situation. In this section we compare theory given in our article and
    in the cited publication by international 
    group and experiment.
    In Conclusion we summarize new results of this article and discuss unsolved problems.
    
    We conclude the introduction by a brief description of the interactions
    essential for helical magnets and the structure of helical ordering
    without defects. In this article we consider systems without inversion
    symmetry. The violation of this symmetry in helical magnets is associated
    with Dzyaloshinskii-Moriya interaction. Typical helical magnets with
    these properties are alloys MnSi \cite{Muhlbauer 2004}, FeGe
    \cite{Uchida 2008}, $\mathrm{Fe_{x}Co_{1-x}Si}$ \cite{Uchida 2006}.
    The Hamiltonian of spin system in these helical magnets in continuous
    approximation can be represented as follows:
    \begin{eqnarray}\label{Hamiltonian}
    \c H=\!\int\! d^3x\left[\frac{J}{2}(\bnabla \bs m)^{2}+g\bs {m}\left(\bnabla\times\bs {m}\right)+\frac{v}{4} \sum_{\alpha=1}^3m_\alpha^4\right].%
    \end{eqnarray}
    Here we measured all lengths in units of the lattice parameter $a$. The first term in (\ref{Hamiltonian}) represents 
    the exchange interaction, the second is the  Dzyaloshinskii-Moriya
    (DM) interaction and 
    the third term is the  crystal field 
    energy corresponding to 
    cubic anisotropy of the lattice. The dimensionless 
    magnetization vector $\mathbf {m}$ in
    this approach has unit length. 
    The hierarchy of interactions is  $J\gg g\gg v$. 
    
    
    In the absence of crystal field 
    the energy of system described by Hamiltonian (\ref{Hamiltonian}) has minimum at magnetization field equal to
    \begin{eqnarray}
    \label{eq: helix}
    \mathbf  m_0(\br)={{\hbe_1}}\cos \mathbf  q\cdot\mathbf  r+{{\hbe_2}} \sin \mathbf  q\cdot\mathbf  r, 
    \end{eqnarray} 
    where $\mathbf  q$ is a vector with fixed modulus $q=q_0\equiv(g/J)\hbq$. 
    The three mutually perpendicular 
    unit vectors $\{\hbe_1, \hbe_2,\hbq\}\equiv\mathbf  T$ form a right tripod  if $g>0$ and a left tripod at $g<0$.  Its orientation in space is arbitrary.  An additional phase in  the argument of sine and cosine 
    can  be absorbed in   a rotation of $\mathbf  T$ around 
    $\hbq$. 
    The only length scale which appears in (\ref{eq: helix}) is  $\ell=2\pi J/g$ which is the pitch of the helix.  In FeGe,
    $\ell  = 70a$. In this structure magnetization is constant in planes perpendicular to the wave
    vector of helix $\mathbf  q$ and rotates at motion along the helix vector.  
    It means that the vectors $\mathbf  q$ and $-\mathbf  q$ correspond to the same
    helical structure. 
    
    Weak cubic anisotropy lifts the 
    degeneracy of the helix energy with respect to direction of $\mathbf  q$. It pins the helix
    axes to the direction of one of the cube diagonals (three-fold axis) if 
    $v > 0$ and
    to the direction of one of 4-fold symmetry axis if $v < 0$. The anisotropy also
    produces little distortions of helical structure which will be neglected. Neutron
    magnetic diffraction experiments \cite{Lebech 1989} have shown that in FeGe the
    axis of helix coincides with the direction (100) or equivalent. It means that the
    constant of cubic anisotropy $v$ in this material is negative.
    
    \section{Topological defects in isotropic helical magnets}
    The central object of the topological classification of defects is
    the degeneracy space $R$ (or order parameter space), i.e. the manifold
    of internal states possessing the same free energy 
    \cite{Toulouse 1976,Volovik 1977,Chaikin 1995,Volovik 2003}.
    The value of order parameter in each point of the $d$-dimensional (sub)space $\mathcal{V}_{d}$
    surrounding the defect determines its mapping  onto the degeneracy
    space $R$. This mapping can be classified into ensembles of equivalent
    maps, which form the $d-$th homotopy group $\pi_d$. If $R$ is disconnected,
    as in systems with a discrete symmetry, then one type of defects are
    domain walls. Inside the wall the order parameter changes between
    its values in the domains, its width is commonly related to the
    strength of the anisotropy responsible for the discrete symmetry.

    To classify defects in helical magnets we  consider first the degeneracy space. 
    Homogeneous rotations of the tripod $\mathbf  T$ induce transitions to other states with the same free energy. Any rotation in 3-dimensional vector space can be parametrized by a vector $\boldsymbol{\omega}$, whose direction defines the axis of rotation and absolute value $\omega$ defines rotation angle. Acting onto a vector $\mathbf  a$ it transforms it into a vector $\mathbf  a'$  defined by equation:
    \begin{eqnarray}
    \nonumber
    \mathbf  a'&=& \mathbf  a \cos\omega\,\mathbf  +(\hbo\times \mathbf  a)\sin\omega+\hbo(\hbo\cdot\mathbf  a)(1-\cos\omega)
    \\\label{eq:rotation}
    &\equiv& e^{\omega\,\hat{\mathbf \Omega}}\mathbf  a=e^{{\boldsymbol \omega} \times}\mathbf  a. 
    \end{eqnarray}
    In the second line  we introduced the linear operator $\hat{\mathbf \Omega}$ defined by its action onto a vector of 3-dimensional space: $\hat{\mathbf \Omega}\,\mathbf  a\equiv\hbo\times\mathbf  a$.
    The rotation angle $\omega$  is confined to the interval  $0\le  \omega\le\pi$. 
    Negative angles  will be  described  as rotations about  $-\hbo$. 
    The set 
    of all rotation vectors $\boldsymbol{\omega}$ fills  
    a 3-dimensional sphere  of the radius $\pi$. 
    Since  rotations around 
    $\hbo$ and  $-\hbo$
    by $\pi$ lead to the same result, 
    diametrically opposite points   $\boldsymbol{t}_\pm=\pm\pi\hbo$ on the  sphere's surface  are equivalent. For equivalence of two operations we will use symbol $\sim$. Thus, $\boldsymbol{t}_+\sim\boldsymbol{t}_-$.  This consideration shows that the order parameter space for isotropic helical magnets is isomorphic to the group of rotation in 3-dimensional space $SO_3$.

    Helix configuration (\ref{eq: helix}) can be considered as a rotation of vector ${\mathbf  e_1}$ about $\hbq$ by an angle $\mathbf  q\cdot\mathbf  r$, i.e.
    \begin{eqnarray}\label{rotation2}
    \mathbf  m_0(\br)=e^{(\mathbf  q\cdot\br)\,\hbq\times }\hbe_1. 
    \end{eqnarray}
    Eq.  (\ref{rotation2}) is  invariant under  the  replacement of $\mathbf  q$ by $-\mathbf  q$. 
    The latter can be treated as a rotation of $\mathbf  q$ around ${\mathbf  e_1}$ by $\pi$. Thus, the invariance with respect to change of sign $\mathbf q$ can be formulated as $e^{\pi\hbe_1\times }\mathbf  m_0=\mathbf  m_0$.   
    In other words,  rotations around ${\mathbf  e_1}$ by $\omega $ and about $-{\mathbf  e_1}$  by $(\pi-\omega)$ are equivalent:
    \begin{eqnarray}
    e^{\omega\hbe_1\times}\mathbf  m_0=e^{-(\pi-\omega)\hbe_1\times}\mathbf  m_0.
    \end{eqnarray}
    Therefore,  also the inner points of the sphere   $\mathbf  u_+=\omega\hbe_1$ and $\mathbf  u_-=(\omega-\pi)\hbe_1$  are equivalent, that is $\mathbf  u_+\sim\mathbf  u_-$. 
    Correspondingly, the order parameter space $R$ is reduced to $R=SO_3/\mathbb Z_2$. This is a sphere of radius $\pi$ with points $\mathbf t_+$ and $\mathbf t_-$, and $\mathbf u_+$ and $\mathbf u_-$ identified.
    Its first  homotopy group  is 
    \begin{equation}\label{Z4}
    \pi_1(SO_3/{\mathbb Z}_2)={
    \mathbb Z}_4,
    \end{equation}
    which is the group of integers modulo 4 \cite{Volovik 1977}.  This result implies that 
    there are three types of topologically stable line defects in helical magnets called disclinations. Analogous defects appear in cholesteric liquid crystals and superfluid $^3$He \cite{Volovik 1977}.
    \\\\
    In disclinations orientation of the tripod $\mathbf  T$  varies in space.  
    Different disclination are characterised by change of the tripod orientation in course of circulation of the coordinate vector $\mathbf r$ along a closed curves  surrounding the central line of defect. These  we parametrize as $\br(s)
    $, where  $s$  denotes a contour variable chaging in the interval $(0,1)$.   For closed curves  $\br(0) =\br(1)$.
    Each closed curve in the real space maps into a closed  curve in the order parameter space $SO_3/\mathbb Z_2$ which we denote as ${\boldsymbol\omega}(s)$. 
    Closed  curves in the order parameter space  can be classified according to their  total rotation angle. 
    \begin{figure}[htb]
        \begin{center}  
            \includegraphics[width=8.3cm]{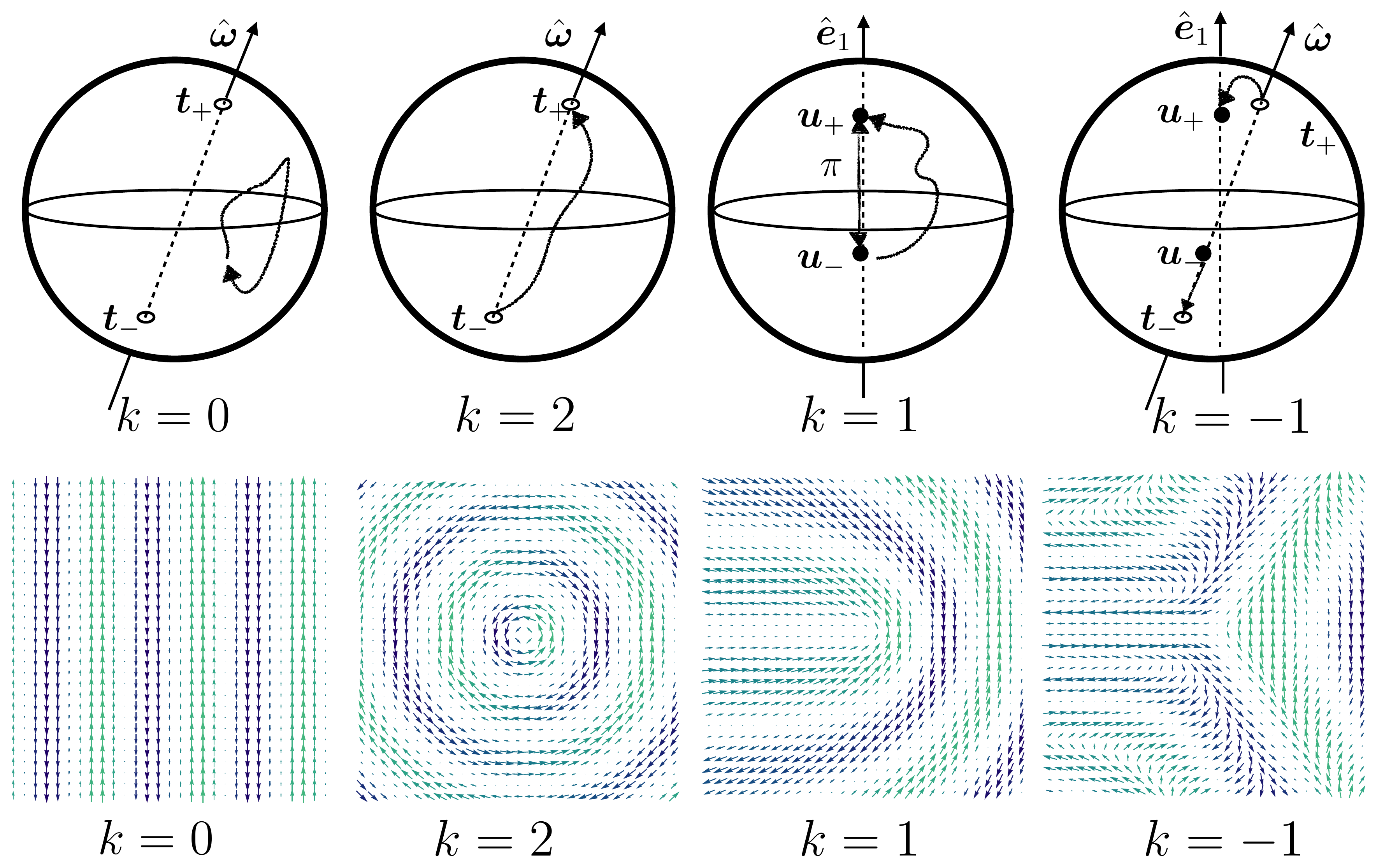}  \caption{Topological defects (disclinations) with  topological  number $k$ in helical magnets. In the upper half different contours are shown as discussed in the text. The lower half shows the corresponding magnetization pattern in the $xy$-plane, as follows from eq.(\ref{magnetization})}\label{Pokrovsky defects}  
        \end{center}
    \end{figure}
    \\
    (i) In the simplest case a closed curve ${\boldsymbol\omega}(s)$ does not contain points ${\mathbf t}_{\pm}$ and ${\mathbf u}_{\pm}$.
    The path can therefore be continuously deformed into a point. The $\boldsymbol \omega$ rotation angle is therefore  zero  (see Fig.\ref{Pokrovsky defects}a). There is no defect
    in this case.\\
    (ii)  
    A   path which  connects the points  ${\mathbf t}_-$ and $\mathbf t_+$.  It corresponds to the rotation angle $2\pi$ (see Fig.\ref{Pokrovsky defects}b). In contrast to the case (i) this contour cannot be deformed into a point. The corresponding defect is called $2\pi$-disclination\\
    (iii)
    Analogously  a  path which  connects points  $\mathbf  u_-$ and $\mathbf u_+$  corresponds to the rotation angle $\pi$ (see Fig.\ref{Pokrovsky defects}c) and corresponds to
    so-called $\pi$-disclination.
    \\
    (iv) A closed curve that connects points ${\mathbf u}_-,{\mathbf u}_+$ in a sequence $\bs  u_-\to\bs t_-\sim\bs t_+\to\bs u_+$
    \\
    (v) More complicated paths which touch equivalent points more than once can be continuously reduced 
    to one of the cases (i)-(iv), as can be seen by explicit construction. A general linear defect in isotropic helical magnet corresponds to rotation to the angle
    $
    \pi [k]_{mod 4}
    $ where $k$ is an integer.
    
    As an example we consider contours $x(s),y(s)$ which include  phase changes  $k\pi$ about  $\hbo\equiv \hbe_1$ by an angle 
    \begin{numcases}
    {\theta[x(s),y(s)]=\frac{k}{2}}\arctan\frac{y}{x} &\textrm {if} $x> 0$\nn\\
    \arctan\frac{y}{x}+\pi&\textrm{if} $x< 0$.\end{numcases}
    $\theta$ has a branch cut at $x=0,y\le 0$.
    The rotated vectors are
    $\hbe_1'=\hbe_1\equiv\hbz$, $\mathbf  q'= \mathbf  q\cos\theta+\hbe_1\times\mathbf  q\sin\theta$ and hence 
    \begin{eqnarray}\label{magnetization}
    \mathbf  m_0'(\br)&=&\hbe\cos\mathbf  q'\cdot\mathbf  r+\hbq'\times\hbe_1\sin\mathbf  q'\cdot\mathbf  r.
    \end{eqnarray}
    The function $\theta(x,y)$ has been used to create the magnetization profile in Fig. 1

    \newcommand{\hbD}{\hat{\b d}}

    \section{Domain walls in the bulk}
    
    Since the anisotropy allows several discrete orientations of the helix,
    wave vector domains with different helix orientation separated by
    domain walls (hDW) can appear. It was shown
    in \cite{Li 2012} that DWs in helical magnets are fundamentally different
    from common Bloch and Neel walls \cite{Bloch 1932,Neel 1948}. These DWs
    are one-dimensional textures in which magnetization rotates around
    a fixed axis. The hDWs are generically two-dimensional textures with
    rotating axis of rotation. For a range of orientations the hDWs contain
    a regular lattice of disclinations. The hDWs that are bisector planes
    between two helix wave vectors in different domains are free of disclinations
    and have minimal surface energy. 
    
    Let the helix wave vectors in two domains be $\mathbf {q}_{1}$ and
    $\mathbf {q}_{2}$. They are not collinear. In the bulk the angles
    between them is either $\pi/2$ if $v<0$ or $\arccos\frac{1}{3}$
    if $v>0$. Since the asymptotic dependence of magnetization in two
    limiting cases cannot be described by one variable, the texture that
    connects the two asymptotical helixes must depend at least on two
    variables. 
    
    The second important fact is that the wave vector inside
    the domain wall necessarily changes its length. Indeed, any continuous
    distribution of magnetization can be described by function $\phi\left(\mathbf {r}\right)$
    whose gradient is the local value of wave vector $\mathbf {q\left(r\right)=\nabla\phi\left(\mathbf {r}\right)}$.
    It is reasonable to assume that $\phi$ depends only on coordinates
    in the plane of the two wave vectors. 
    In other words, the domain wall
    must be a plane perpendicular to the plane $\left(\mathbf {q}_{1},\mathbf {q}_{2}\right)$.
    Let us assume that the lines of wave vectors are continuous curves
    asymptotically approaching straight lines in direction $\mathbf {q}_{1}$
    and $\mathbf {q}_{2}$ in different domains. Magnetization is described
    by eq. (\ref{eq: helix}) in which the argument of sine and cosine is replaced by $\phi(\mathbf{r})$. It is possible to take unit vector
    $\hat{\mathbf e}_{1}$ perpendicular to the plane $\left(\mathbf {q}_{1},\mathbf{q}_{2}\right)$.
    The local vector $\hat{\mathbf{e}}_{2}\left(\mathbf{r}\right)$ is uniquely
    determined by the vectors $\mathbf{q}\left(\mathbf{r}\right)$ and
    $\hat{\mathbf{e}}_{1}$. The requirement of constant modulus for wave vector
    leads to equation $\left(\nabla\phi\right)^{2}=\mathrm{const}$. This
    equation coincides with the stationary Hamilton-Jacobi equation for
    free particle. The vector $\nabla\phi$ is the momentum of this particle.
    But free particle can not change its momentum. Therefore, there is
    no solution of such equation with asymptotics of $\nabla\phi$ equal
    to one constant vector in one domain and another constant vector in
    another domain. 
    
    Another consequence of this consideration is that the width of the
    hDW has the order of magnitude of the pitch of helix $\ell$. Indeed,
    the change of wave vector modulus violates the balance of exchange
    and DM forces. Namely they define the variation of magnetization inside
    the domain wall and restore theis balance outside. The contribution
    of anisotropy can be neglected. The only scale of the isotropic Hamiltonian
    is $\ell$. This peculiarity of the hDW is also unusual. Commonly the
    width of domain wall is determined by competition of exchange force
    and anisotropy. Such a domain wall would be much wider. 
    
    Consider the domain wall that is a bisector plane between vectors
    $\mathbf{q}_{1}$ and $\mathbf{q}_{2}$. Its normal  is directed
    along the unit vector $\hbq_- $ where 
    \begin{equation}
    \mathbf q_\pm=\frac{1}{2}\lk\mathbf q_1\pm\mathbf q_2\rk
    \end{equation}
    Both vectors $\mathbf{q}_{+}$ and $\mathbf{q}_{-}$ lay in the plane perpendicular to the domain wall. 
    Let assume that vector field $\mathbf{q}\left(\mathbf{r}\right)$
    asymptotically approaches $\mathbf{q}_{1}$ in the domain $\hat{\mathbf q}_-\cdot \bs r>0$
    and $\mathbf{q}_{2}$ in the domain $\hat{\bs q}_-\cdot \bs r<0$. 
    
    A simple non-singular trial function
    for the phase $\phi\left(\mathbf{r}\right)$ in the presence of domain wall has
    a form:
    \begin{equation}
    \phi\left(\mathbf{r}\right)=\mathbf{q}_{+}\mathbf{r}+q_{-}w\ln\left[2\cosh\left(\frac{\mathbf{\hbq_-\cdot r}}{w}\right)\right]\label{eq:bisector-trial}
    \end{equation}
    The vector field $\mathbf{q}\left(\mathbf{r}\right)$ for this trial
    function reads:
    \begin{equation}
    \mathbf{q\left(r\right)=}\mathbf{q}_{+}+\mathbf{q}_{-}\tanh\left(\frac{\mathbf{\hbq_-\cdot \br}}{w}\right)\label{eq:q}
    \end{equation}
    The magnetization in such texture can be represented as follows:
    \begin{equation}
    \mathbf{m\left(\mathit{x,y}\right)=}\hat{\mathbf z}\cos\phi\left(\mathbf{\mathit{x,y}}\right)+\hat{\mathbf q}\times\hat{\mathbf z}\sin\phi\left(\mathbf{\mathit{x,y}}\right)\label{eq:m}
    \end{equation}
    Here $\hat{\mathbf z}$ is unit vector in the direction of vector $\mathbf{q}_{1}\times\mathbf{q}_{2}$,
    $\hat{\mathbf q}$ is unit vector in direction of $\mathbf{q\left(r\right)}$;
    $x-$axis is parallel to the vector $\mathbf{q}_{+}$. In Fig. 2 the
    regions of positive and negative projections of $\mathbf{m\left(r\right)}$
    to the $z-$axis and local direction of $\mathbf{q\left(r\right)}$
    are schematically shown for $w=0.3\ell$. Minimization of energy (see section 4) for
    mutually perpendicular $\mathbf{q}_{1}$ and $\mathbf{q}_{2}$ gives
    $w\approx 0.3\ell $. In order to make magnetic texture
inside the hDW clearly visible we display in Fig. 2 a thicker domain wall.
    \begin{figure}[htb]
        \begin{center}  
            \includegraphics[width=6.3cm]{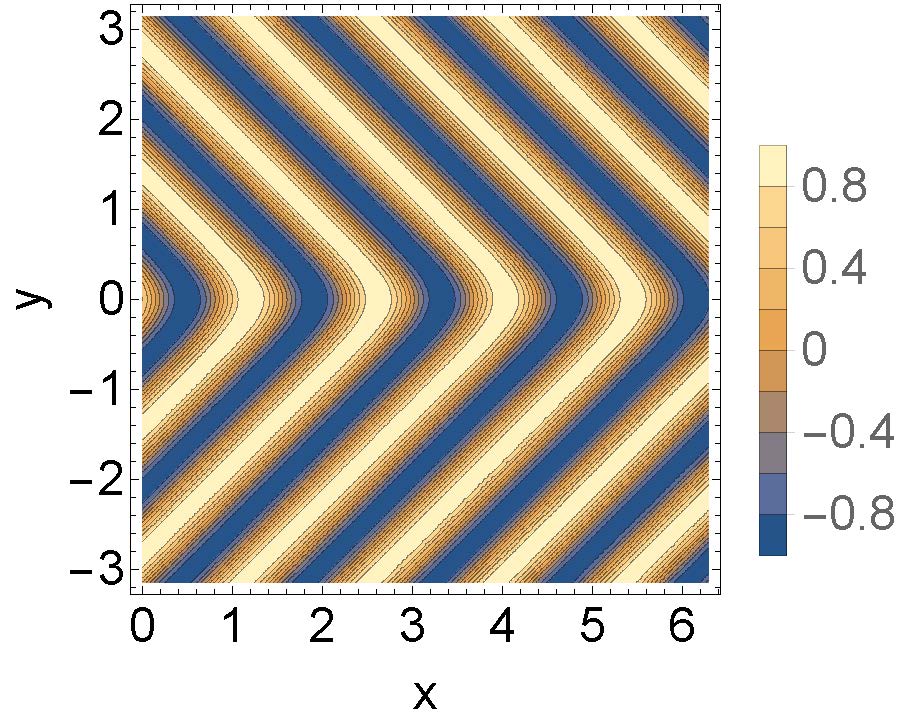}\caption{Distribution of the magnetization projection $m_{z}$ in the bisector domain wall according to trial function (\ref{eq:bisector-trial}) with $w=0.3\ell$. The value of $m_z$ is shown by color} \label{H:bisector-DW}  
        \end{center}
    \end{figure}
    \\
    Note that there are two bisectors for any pair of wave
    vectors in domains. Their trial functions differ by permutation of
    vectors $\mathbf{q}_{+}$ and $\mathbf{q}_{-}$. They both realize
    local angular minima of the surface energy, but their energies per
    unit area are different. Namely, the bisector of the angle $\alpha$
    between wave vectors less than $\frac{\pi}{2}$ has smaller surface
    energy than bisector of complementary angle $\pi-\alpha$ (see Fig.\ref{wall types}). 
    \begin{figure}[htb]
        \begin{center}  
            \includegraphics[width=8.3cm]{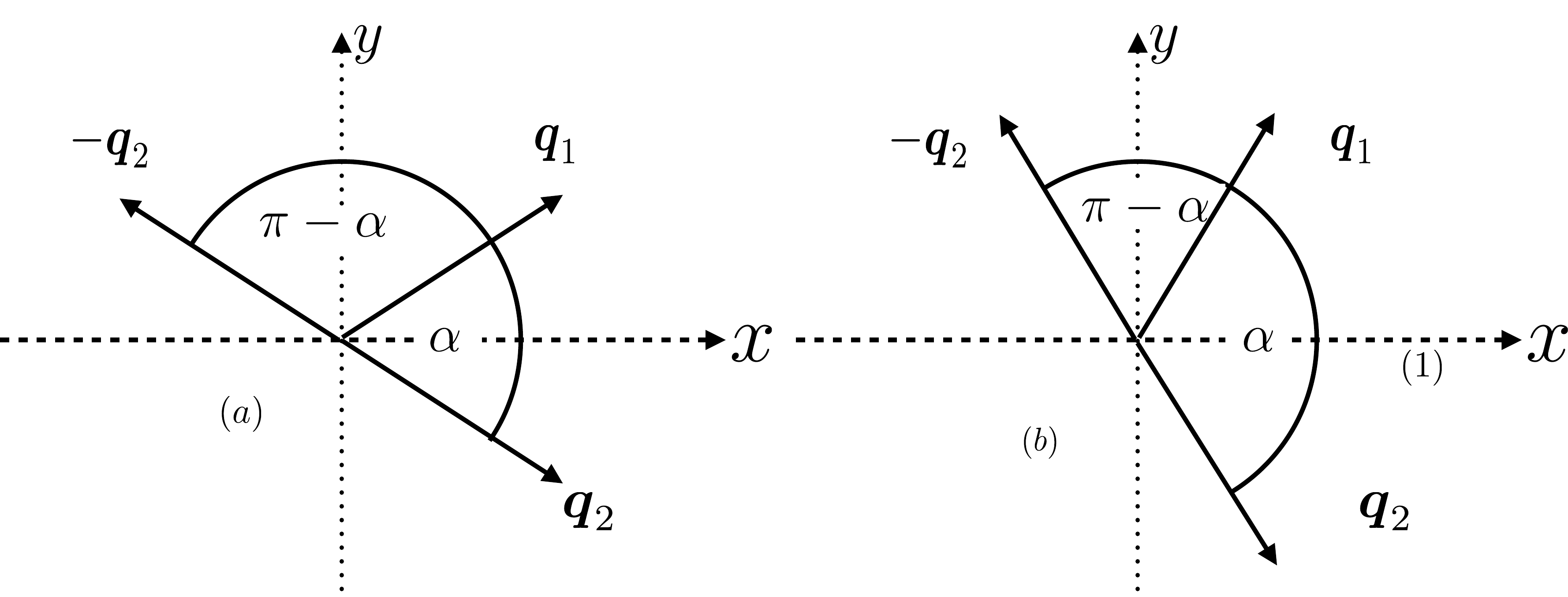}  \caption{Wave vectors $\b q_1$ and $\b q_2$ and bisector domain walls (dashed or dotted).}\label{wall types}  
        \end{center}
    \end{figure}
    Any attempt 
    to extend such Ansatz to the case of domain wall of other
    orientation leads to divergent field $\mathbf{q\left(r\right)}$.
    The reason of that can be illustrated by example of mutually perpendicular
    $\mathbf{q}_{1}$ and $\mathbf{q}_{2}$ and the domain wall perpendicular
    to $\mathbf{q}_{1}$ and parallel to $\mathbf{q}_{2}$. Then in the
    plane of domain wall at approaching from the $\mathbf{q}_{1}$-domain,
    magnetization $\mathbf{m}$ is constant that means constant $\phi$,
    while at approaching from  $\mathbf{q}_{2}$-domain, the phase
    is linear function of coordinates in plane and goes to infinity when
    coordinate $x$ goes to infinity. Since the transition from one asymptotic
    to another proceeds in the interval of finite width $\sim \ell$, $\left|\nabla\phi\right|$
    also goes to infinity. This enormous mismatch can be avoided only
    by singularities that interrupt continuous counting of phase and allow
    to keep $\mathbf{q\left(r\right)}$ in finite limits. In our work
    \cite{Li 2012} we assumed that these compensating singularities are
    vortices or in  terminology accepted for isotropic helical magnets (see section 2) $2\pi$-disclinations. 
    However, as it was first understood by M. Garst \cite{Garst 2015}
    the $\pm\pi$-disclinations can do the same job with smaller energy
    price. Strictly speaking, disclinations are forbidden by discrete
    symmetry. However, they still can exist on a small length scale if
    their fields are compensated on distances much less than characteristic
    length at which the anisotropy becomes essential $\ell_{an}\sim\sqrt{J/v}a\sim \ell\frac{J}{g}$,
    i.e. in the range of 100 times larger than pitch.
    \begin{figure}[htb]
        \begin{center}  
            \includegraphics[width=5.3cm]{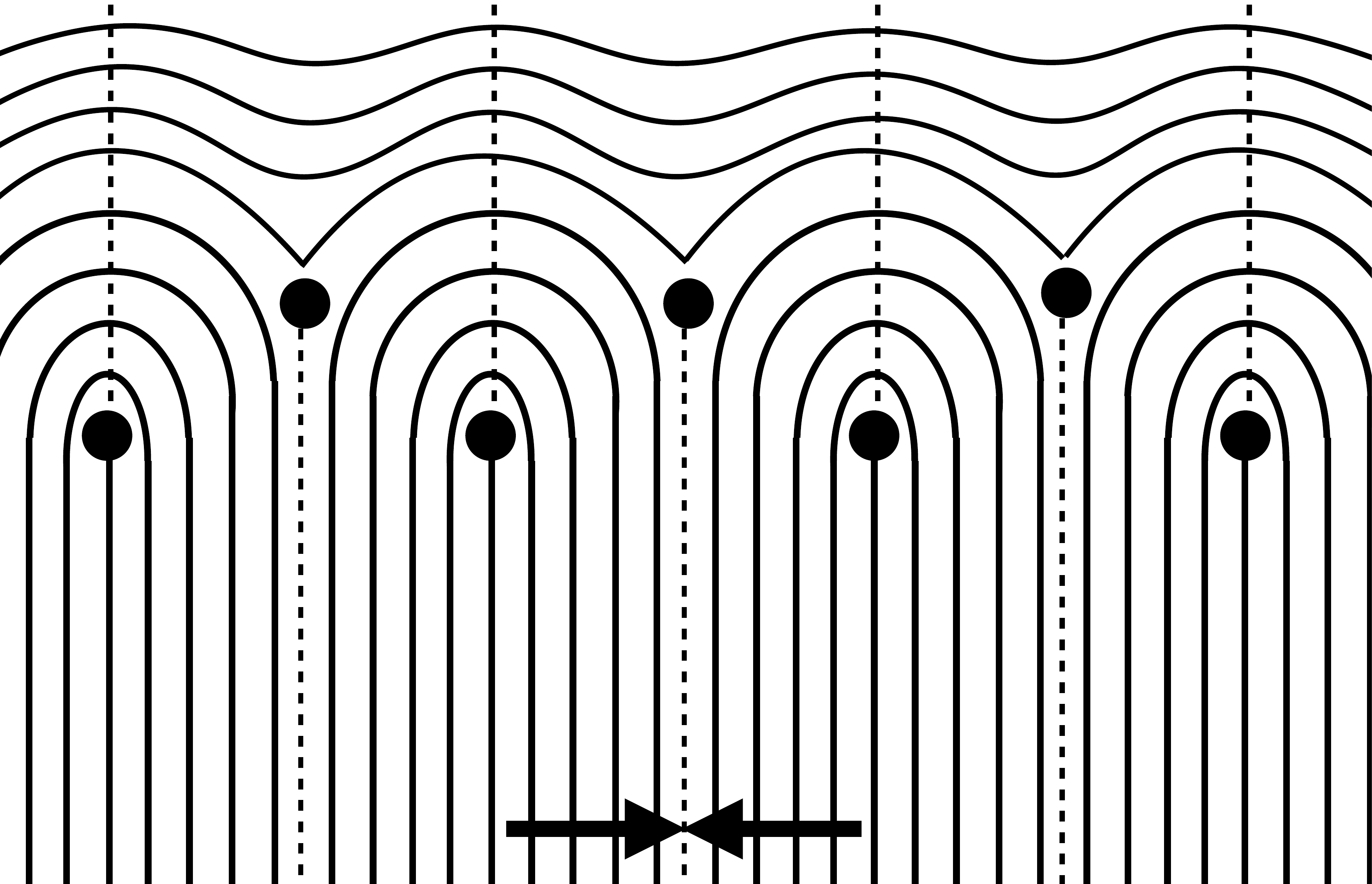}  \caption{Schematic plot of zig-zag domain wall formed by alternating $\pm\pi$-disclinations. The wave vectors near a separatrix are shown by arrows.}\label{H:zizag}  
        \end{center}
    \end{figure}
    As it is clear from schematic picture Fig. 4 a periodic chain of $\pm\pi$-disclinations
    forms a zig-zag-shaped domain wall separating regions with mutually
    perpendicular wave vectors (one of them is parallel to the domain
    wall). In the work \cite{Li 2012} we predicted appearance of zig-zag
    domain walls. However, as the reason for zig-zag shape we indicated
    the instability of a domain wall whose orientation deviates from bisectors
    with respect to formation of a zig-zag whose sides are pieces of vortex-free
    domain walls, i.e. bisectors. Then two directions of zig-zag lines
    must be mutually perpendicular, whereas in the case of dominating
    disclination-anti-disclination structure the angle between them must
    be 120$^\circ$. Zig-zag structure was first observed in experiments with
    alloy Fe$_{0.5}$Co$_{0.5}$Si.\cite{Uchida 2006} Fluctuations of
    the direction in these experiments were large and it is difficult
    to say what angle dominates. Topologically the vertex lines of zig-zag
    hDW are discilnations independently on what mechanism dominates. If
    the sides of zig-zag are large in comparison to pitch, both nechanisms
    can work at different scales. The most important new discovery made
    by the international team \cite{Schoenherr 2018} missed in our work
    \cite{Li 2012} is that the vertices 
    of zig-zag are simultaneously centers of disclinations.
    
    Let consider zig-zag structure in more details. It is possible to
    separate it in primitive cells each containing one $+\pi$ and one
    $-\pi$ disclinations. Each disclination has its own part of primitive cell 
    either above upper vertex or below lower vertex of zig-zag. The parts belonging 
    to different disclinations are separated by separatrices shown in Fig. 4 by dashed 
    straight lines. Let us consider now
    two neighboring lower disclinations. Note that the angle between the
    wave vector and domain wall median plane ($x-$axis) near lower separatrix
    far from domain wall is equal to 0 for disclination left and $\pi$ for disclination
    right of separatrix. This seeming discontinuity in reality does not
    exist since these directions of wave vector are equivalent and magnetization
    remains continuous at separatrix. At a separatrix the phase $\phi$ experiences
    jump by $2\pi n$ with some integer $n$ that reduces it to the basic interval $(-\pi n,\pi n)$, but magnetization
    remains continuous.  This is just the mechanism of mismatch
    compensation. 
    In the range between upper and lower zig-zag vertices, there is no
    clear separation between the lines related to upper and lower disclinations.
    If the sides of zig-zag are significantly longer than the pitch, then
    the zig-zag sides can be considered as pieces of domain walls. 
    
    Let us calculate roughly the energy of the disclination hDW. With
    this purpose we will derive exact general equation for the domain
    wall energy as functional of the wave vector field $\mathbf{q\left(r\right)}$
    that will be useful for other problems. Exchange energy is proportional
    to the square of tensor $\nabla\mathbf{m}$. According to eq. (12)
    components of this tensor read:
    \begin{equation}
    \partial_{\alpha}\mathbf{m}=q_{\alpha}\hat{\mathbf q}\times\mathbf{m}+\left(\partial_{\alpha}\hat{\mathbf q}\right)\times\hat{\mathbf z}\sin\phi\label{eq:grad-m}
    \end{equation}
    Vector $\partial_{\alpha}\hat{\mathbf q}$ is orthogonal to $\hat{\mathbf q}$. Therefore
    it is a linear combination $\partial_{\alpha}\hat{\mathbf q}=a\hat{\mathbf z}+b\hat{\mathbf q}\times\hat{\mathbf z}$.
    Since vector $\hbq$ at any $\mathbf{r}$ lies in the $x,y$-plane, coefficient $a$ is zero.
Thus $\partial_{\alpha}\hat{\mathbf{q}}\times\mathbf{z}=-b\hat{\mathbf q}$. It means that the first and second terms
    in the right hand  side of eq. (\ref{eq:grad-m}) are mutually orthogonal.
    Thus we find:
    \begin{equation}
    \left(\nabla\mathbf{m}\right)^{2}=\mathbf{q}^{2}+\sin^2\phi\left[\left(\partial_{\alpha}\hat{\mathbf q}\right)\times\hat{\mathbf z}\right]^{2}\label{grad-m-square}
    \end{equation}
    Since $\partial_{\alpha}\hbq$ is a vector perpendicular to $\hat{z}$, the expression $[\partial_{\alpha}\mathbf{q}\times z]^2$ can be replaced by $(\partial_{\alpha}\hbq)^2$. To calculate the DM energy we need to find curl of magnetization:
    \begin{equation}
    \nabla\times\mathbf{m}=-q\left(\mathbf{r}\right)\mathbf{m}+\hat{\mathbf z}\left(\nabla\hat{\mathbf q}\right)\sin\phi\label{curl-m}
    \end{equation}
    and 
    \begin{equation}
    \mathbf{m\cdot\nabla\times\mathbf{m}}=-q\left(\mathbf{r}\right)+\left(\nabla\hat{\mathbf q}\right)\sin\phi\cos\phi\label{eq:m-curl-projection}
    \end{equation}
    Finally the total domain wall energy per unit area reads:
    \begin{equation}
    E_{dw}=\frac{J}{2L}\int_{0}^{L}dx\int_{-\infty}^{\infty}dy\left\{ \left[q\left(\mathbf{r}\right)-q_{0}\right]^{2}+\frac{1}{2}\left(\partial_{\alpha}\hat{\mathbf q}\right)^2\right\}, \label{eq:dw-energy}
    \end{equation}
    where $L $ is period of zig-zag structure. The energy does not change if the phase 𝜙 in eq. (12) for
magnetization changes by any constant. Therefore, in expression for energy the factor
sin> 𝜙 of eq. (14) can be replaced by ½ and the second term of eq. (16) vanishes. 
 Eq. (\ref{eq:dw-energy}) clearly demonstrates
    that domain wall energy consists of two independent parts. One of
    them is due to deviation of modulus of wave vector from its most energy
    favorable value $q_{0}$, whereas the second is due to change of its
    direction. 
    
    Applying this general equation to disclination domain wall, we conclude
    that square of deviation $\left[q\left(\mathbf{r}\right)-q_{0}\right]^{2}$
    has order of magnitude $q_{0}^{2}$ inside domain wall and quickly
    decreases outside. Therefore integral from this term by order of magnitude
    is equal to $L^{2}q_{0}^{2}$. The second term in curl brackets due
    to rotation of wave vector is proportional to $1/r^{2}$ near each
    center of disclination, where $r$ is the distance to the center.
    After integration it gives $\ln\frac{L}{\ell}$ by order of magnitude.
    Thus, the total energy of disclination hDW per unit area is roughly
    equal to
    \begin{equation}
    E_{dw}\sim Jq_{0}^{2}L+J\frac{\ln\frac{L}{\ell}}{L}\label{eq:DW-rough}
    \end{equation}
    Its minimization gives $L\sim q_{0}^{-1}\sim\ell$. More accurate
    coefficients in this relation can be found numerically. It requires
    sufficiently accurate solution of static Landau-Lifshitz equation
    with singularities or a proper trial function for $\phi(x,y)$ that we did not find so far.
    
    \section{Domain walls at crystal faces}
    
    In the cited work \cite{Schoenherr 2018} the authors studied about
    90 samples of FeGe single crystals with typical sizes 0.5$\times$1$\times$1
    mm. They were prepared by K. Nakamura under supervison of  Y. Tokura  by very slow (1 month) growth from vapors  in vacuum quartz tube at temperature bias between 500
    and 560$^\circ$ C. The crystal structure was checked by Laue diffraction. The
    samples then were cut and polished to yield faces (100) and (110)
    with roughness 1 nm. MFM pictures and measurements were performed in Trondheim by P. Schoenherr under supervision of D. Meier.  The magnetic tip in the MFM had radius about
    50 nm. It was scanned with the distance of the tip to face 30 nm.
    Standard dual-pass MFM measurements allowed the resolution 10-15 nm. Measurements
    were performed at temperature 260-273K maintained by permanent water flux. M. Garst and A. Rosch supervised theoretical part of work.
    
    The first experimental fact discovered in the MFM studies of helical
    magnet \cite{Schoenherr 2018} is that on the crystal face, the wave
    vectors of helix lay in the plane of face. That was checked for the
    faces (1,0,0) and (1,1,0). The second surprising fact is that unlike
    in the bulk, the orientation of the surface helix wave vector is not
    confined to a definite crystallographic directions within the face
    plane. These two facts can be explained if spin-orbit interaction
    near the face creates uniaxial easy-axis anisotropy. Let us denote $\mathbf{n}$  the 
    normal vector to the face. Then surface anisotropy energy is 
    \begin{equation}
    H_{sa}=-\lambda\int\mathbf{\left[n\cdot m\left(r\right)\right]}^{2}\frac{d^{2}x}{a^{2}}\label{eq:sur-an}
    \end{equation}
    The unit magnetization vector field is given by eq. (\ref{eq:m}).
    Let choose $\mathbf{e}_{1}=\frac{\mathbf{n\times}\mathbf{\hat{\mathbf q}}}{\left|\mathbf{n\times}\mathbf{\hat{\mathbf q}}\right|}$
    and $\mathbf{e}_{2}=\mathbf{\hat{\mathbf q}\mathbf{\times e}_{1}}=\frac{\mathbf{n}-\hat{\mathbf{q}}\left(\mathbf{n\hat{\mathbf q}}\right)}{\left|\mathbf{n\times}\mathbf{\hat{\mathbf q}}\right|}$.
    Then $\mathbf{nm\left(r\right)}=\mathbf{ne_{2}\sin\phi\left(\mathbf{r}\right)}=\left|\mathbf{n\times}\mathbf{\hat{\mathbf q}}\right|\sin\phi\left(\mathbf{r}\right)$.
    Average $\left\langle \mathbf{\left[n\cdot m\left(r\right)\right]}^{2}\right\rangle $
    over phase $\phi$ is equal to $\frac{1}{2}\left(\mathbf{n\times}\mathbf{\hat{\mathbf q}}\right)^{2}=\sin^{2}\theta$
    where $\theta$ is the angle between helix wave vector $\mathbf{q}$
    and the  normal to the face $\mathbf{n}$. Thus, the surface anisotropy
    energy per unit area is $\sigma=-\frac{\lambda}{2}\sin^{2}\theta$.
    It has minimum at $\theta=\frac{\pi}{2}$, i.e. for the helix wave
    vector in the plane of face. The value of surface anisotropy $\lambda$
    appears as relativistic correction of the second order, whereas the
    constant of cubic anisotropy $v$ appears only in the fourth order
    relativistic correction. Therefore it is reasonable to assume that
    $\lambda\gg v$. This fact explains why the helix wave vector tends
    to turn parallel to the  face plane. The in-face anisotropy 
    is too weak to confine these vectors to crystal directions
    with small indices in the face. In infinite perfect samples the helix
    vector on surface is determined by its bulk value and the normal to
    the face with some exceptions. For the face (001) and the same helix
    wave vector in the bulk, any direction of surface helix wave vector
    has the same energy. The authors of article \cite{Schoenherr 2018}
    concluded that there is no dependence between bulk and surface wave
    vectors. When the bulk vector is not perpendicular to the crystal
    face as happens for the face (110), this result looks surprising.
    It may happen if the parameters $J$ and $g$ 
    rapidly change in close
    vicinity of the boundary. Then surface layer becomes to some extent
    magnetically independent of the bulk. 
    An alternative explanation proposed 
    in \cite{Schoenherr 2018} is the closeness of temperature 260-273 K at which 
    measurements were performed to the Neel point $T_N=278 K$. In this range of temperature the magnetization is still small and energy of bulk anisotropy proportional to the fourth
    power of magnetization becomes negligible in comparison to other contribution to energy quadratic in $\mathbf{m}$. 
    
    Due to insensitivity of energy to the direction of surface helix
    wave vectors, the angle $\alpha$ between wave vectors $\mathbf{q}_{1}$
    and $\mathbf{q}_{2}$ in different domains changes from sample to
    sample. Symmetry of wave vectors with respect to change of sign implies
    that $\alpha$ varies in the limits between 0 and $\pi$. Let us denote
    $\beta$ the angle formed by one of the helix wave vectors and domain
    wall. Experimental graph of dependence of $\beta$ on $\alpha$ is
    shown in Fig. 5d.
    \begin{figure}[htb]
        \begin{center}  
            \includegraphics[width=7.3cm]{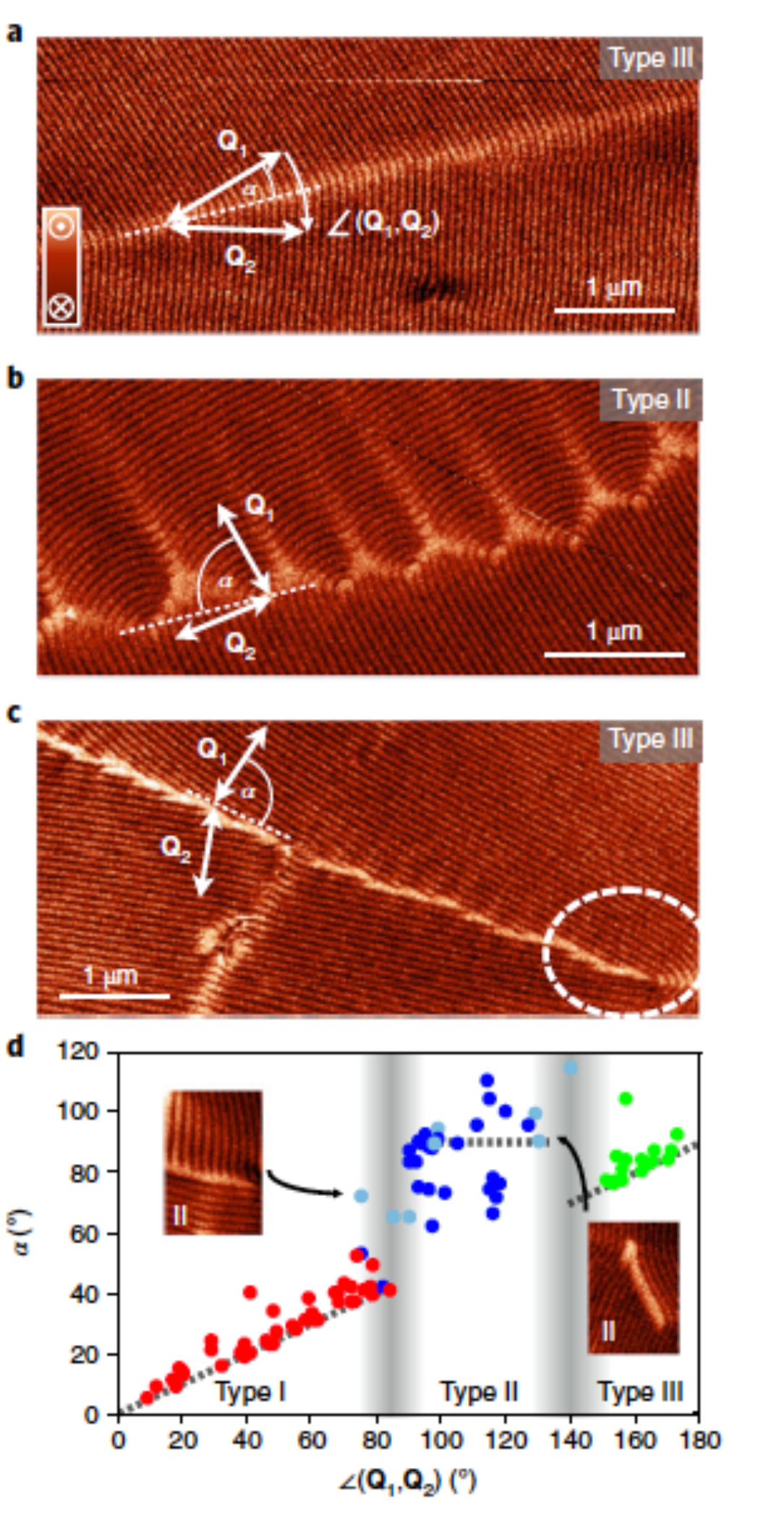}  \caption{MFM pictures of Helimagnetic domain walls in FeGe. Reproduced from
                article \cite{Schoenherr 2018} by courtesy of Profs.  D. Meier and Y. Tokura. a. Bisector domain wall at $\alpha<80^{\circ}$ (type I); b. Zig-zag disclination wall (type III);c. Bisector domain wall at $\alpha>140^{\circ}$ (dislocation wall, type II); d. Graph of the angle $\beta$ between domain wall and one of the helix wave vectors (our notations differ from those accepted in \cite{Schoenherr 2018}). See comments in the text.}  
            \label{H:experiment}  
        \end{center}
       \end {figure} 
        \\
        It shows that at $0<\alpha<80$$^\circ$ (DW of the type
        I in terminology accepted in \cite{Schoenherr 2018}) and at 140$^\circ$$<\alpha<180$$^\circ$
        (DW of type III), $\beta$ follows well defined dependence $\beta=\alpha/2.$
        These are domain walls whose plane are bisectors between vectors $\mathbf{q}_{1}$
        and $\mathbf{q}_{2}$ that are well described by variational eqs.
        (\ref{eq:bisector-trial},\ref{eq:q}). Experimental data imply that
        domain walls in this ranges of $\alpha$ relax to the closest bisector
        direction. Note that the angle between any possible initial direction
        of the hDW and bisector in this range of angles is less than 40$^\circ$.
        In the interval 140$^\circ$$<\alpha<180$$^\circ$ (DW of the type II), $\beta$
        is not a function of $\alpha.$ Experimental values of $\beta$ at
        fixed $\alpha$ are scattered in this range of $\alpha$ more or less
        uniformly between $\alpha/2$ and $\alpha$. Such domain walls
        must be supplied with zig-zag chain of disclinations-antidisclinations
        as discussed in Section 3. However, at $\alpha=90$$^\circ$ and a fixed value
        $\beta\neq0,\frac{\pi}{2}$, the two lines within a primitive period
        have different lengths. It follows from geometrical constraints, i.e.
        fixed values $\alpha$ and $\beta$ and angles (120$^\circ$ or 90$^\circ$) at vertices
        of zig-zag line. MFM pictures of the domain walls in this range of
        $\alpha$ confirm the existence of disclination-antidisclination zig-zag
        structure, though in real crystals it is not so ideally periodic and
        domain wall median is not ideal straight line
        
        In theoretical part of the article \cite{Schoenherr 2018}, the authors
        performed micromagnetic calculations of domain wall configuration
        in two dimensions. They demonstrated that at small and large angles
        $\alpha$, minimum energy is realized by smooth non-singular bisector
        domain walls. At angles $\alpha$ in the range near 90$^\circ$, the domain
        walls has zig-zag shape with regularly intermitting disclinations
        and anti-disclinations. Theory even describes the irregularity of this
        chain considering them as random fluctuations. This is without doubt
        a success. However some principal questions remain unanswered. 
        
        The main such question is what keeps orientation of the helix wave vectors?
        We understand that in the bulk it is anisotropy that reduces initial
        SO(3) symmetry of the exchange and DM interactions to discrete group
        of cube without inversion. But experimenters tell us that anisotropy
        plays no role at the crystal faces. Then there is no topological reason
        for appearance of domain walls.
        Another idea is that the helix on the surface is fixed by its coupling
        with the helix in the bulk. In this case orientation of the wave vector
        at the surface could be arbitrary in plane of the face (001), but
        not in the face (110). It also may be the edge or random pinning that fixes the
        orientation of wave vector near the boundaries or defects and then
        these fixed pieces serve as nuclei for domain growth.  
        
        But if the direction of helix vector can vary continuously another principal 
        question appears. If symmetry of a system does
        not require domain walls as topological defects, can they nevertheless appear if different
        possible values of order parameter are fixed 'by hands'' near boundaries of the sample?
        For simple systems such as Heisenberg or XY (planar) magnets the answer
        to this question is no. In both cases, if spins on two sides of a
        big stripe or slab are fixed artificially, the transition from one
        to another orientation in the sample proceeds smoothly. Instead of
        domain wall we see spins slowly rotating in space. The principal question
        is whether the same is correct for an isotropic helical magnets. The
        difference with simpler systems is that there is no smooth texture
        that changes its direction conserving the modulus of wave vector (see
        Section 3). In this article we issue a proof of the statement that in isotropic
        helical magnet, the transition between two different helix wave vectors fixed
        near boundaries proceeds by formation of the hDW whose width depends on
        the angle $\alpha$ between fixed wave vectors. We calculate explicitly the angular dependence
        of the width for bisector domain walls given by variational ansatz (\ref{eq:bisector-trial}). 
        
        The derivation of this result employs general equation (\ref{eq:dw-energy})
        with $\mathbf{q}(\mathbf{r})$ given by Ansatz (\ref{eq:q}). It represents the domain wall energy
        as a sum of two positive contributions $E_{dw}^{(1)}$ and $E_{dw}^{(2)}$ originating from deviation of $q(\mathbf{r})$
        from its energy preferable value $q_0$ and from rotation of the wave vector $\hat{\mathbf q}(\mathbf{r})$, respectively. The integrals involved
        in these expressions can be calculated explicitly. The result is as follows:
        \begin{equation}\label{eq:bisector-dev}
        E_{dw}^{(1)}=\frac{1}{2}Jq_0^{2}wI_1,
        \end{equation}
        where
        \begin{eqnarray}\label{eq:I1}
        I_1(\alpha)&=&\frac{\sin\alpha}{2}\lk-\frac{a^{2}-1}{a}+A\ln a+B\ln\frac{2a}{a^2+1}\rk
        \end{eqnarray}
        and 
        \begin{eqnarray}\label{eq:aBCD}
        a&=&\cot\frac{\pi -\alpha}{4},\,\,A=\frac{2\left(a^{2}+1\right)}{a},\,\,B=\frac{2\left(a^{2}+1\right)^{2}}{a\left(a^{2}-1\right)}.
        \end{eqnarray} 
        The rotational part of energy reads $E_{dw}^{(2)}=\frac{J}{w}I_{2}$
        where 
        \begin{equation}\label{eq:bisector-rot}
        I_{2}(\alpha)=\frac{1}{2}\left( \alpha -
\cot\alpha\right)
        \end{equation}
        Minimization of  the total DW energy $E_{dw}=E_{dw}^{(1)}+E_{dw}^{(2)}$ over $w$ leads to the result:
        \begin{equation}\label{eq:width}
        w(\alpha)=q_0^{-1}\sqrt{\frac{I_2}{I_1}}
        \end{equation}
        This equation shows that the width of bisector domain wall becomes infinite at $\alpha=0$, { has a minimum  and again goes to infinity at $\alpha=\pi$. The curve  $w(\alpha)$ (see Fig. 5) is not symmetric with respect to the point $\alpha=\pi/2$, i.e. $w(\alpha)\neq w(\pi - \alpha )$. This asymmetry seemingly contradicts to the established invariance of the helix to the flip of one of wave vectors. The reason of this  asymmetry is the accepted assumption that  domain wall is the bisector of a smaller angle between wave vectors entire interval $0<\alpha <\pi$. Certainly, the second half of this interval can be reduced to the first by the flip, for definiteness, of $\mathbf{q}_2$, i.e. by transformation $\alpha\leftrightarrow \pi - \alpha $. However, at this flip the domain wall that forms an angle $\pm(\pi-\alpha)/2$ with wave vectors is bisector of not the smaller, but larger angle between two axis of rotations (see Fig. 3). 
            
            More formally this fact can be formulated as follows: at accepted assumption, the symmetry transformation is $\alpha\leftrightarrow\pi - \alpha $ and simultaneously $\mathbf{q}_+\leftrightarrow\mathbf{q}_-$. As we stated earlier, the surface energy of the bisector of larger angle hDW is larger than the analogous energy for smaller angle. Note that both are local minima of the surface energy $\sigma$ as function of the angle $\beta$ between plane of domain wall and one of the wave vectors at fixed value of $\alpha$.} 
        In the bulk the angle between different wave vectors is $\alpha=\frac{\pi}{2}$. Then $a=\sqrt{2}+1\approx 2.414$ and the domain wall width of
        bisector domain wall is $w\left(\pi/2\right) \approx 0.038\ell$. 
        \begin{figure}[htb]
            \begin{center}  
                \includegraphics[width=6.3cm]{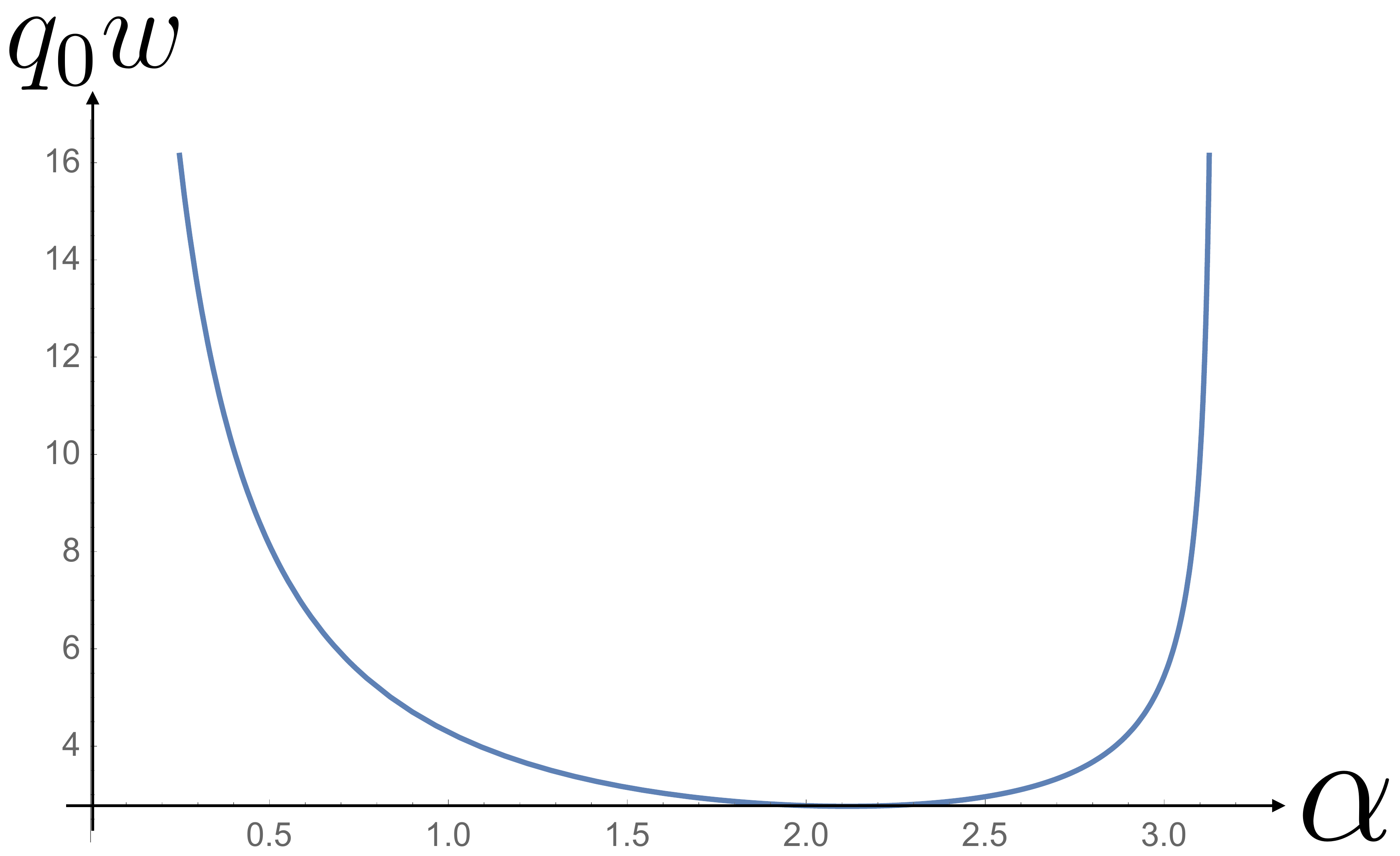}  
                \caption{Graph of function $\sqrt{2}q_0w(\alpha)$ for the width of bisector hDW. }\label{H:w-alpha}  
            \end{center}
        \end{figure}
        
        Though our calculation relates to a specific choice of trial function, the main features of the function $w(\alpha)$ minimizing the surface energy 
        remains the same for exact solution of this problem. Namely, $w(\alpha)\rightarrow\infty$ at $\alpha\rightarrow 0$ and $\alpha\rightarrow \pi$; it has minimum  and it  goes to infinity at $\alpha=\pi/2$. It is asymmetric with respect to reflection in the point $\alpha=\pi/2$. At this value of $\alpha$, both angles between axis of rotations are equal. Therefore, in this case the width and the energy of two branches of the curves $w(a)$ and $E(\alpha)$ arrive at the common limit. } 
    
    { We start the proof of these statements with the case of small $\alpha$. In general case the wave vector $\mathbf{q}(x,y)$ must go to the limit $\hat{x}q_0$ at $\alpha\rightarrow 0$. Corrections to this limit at small $\alpha$ must start at least with $\alpha^2$ since at permutation of two wave vectors  the distribution of magnetization does not change. It means that modulus of wave vector $q(x,y)$ differs from $q_0$ by a small value of the relative order  $\alpha^2$. As a consequence $\vert q(x,y)- q_0\vert^2$ has order of magnitude $\alpha^4q_0^2$ and $I_1\sim \alpha^4$.  The unit vector $\hat{\mathbf{q}}(x,y)$ rotates in the domain wall to the angle $\alpha$. Therefore, $\nabla\hat{\mathbf{q}}$ is $\sim\alpha/w$ by the order of magnitude. Thus, $ I_2\sim\alpha^2$.
        We conclude that $w(\alpha)\propto 1/\alpha$ at $\alpha\rightarrow 0$. 
        
        At $\alpha$ approaching $\pi$,  the $x-$component of wave vector tends to zero. Neglecting it, we arrive at wave vector that has only $y-$ component. Due to symmetry, it changes sign at crossing the median of hDW $y=0$. Therefore, the unit vector of its direction is equall to sign of $y$ and its derivative $\nabla\hat{\mathbf{q}}=2\delta(y)$. Thus, $I_2=w\int_0^{\infty} (\nabla\hat{\mathbf{q}})^2d\eta$ diverges at $\alpha$ approaching $\pi$. On the other hand, integral $I_1$ remains finite in this limit since $q$ is even function of $y$ that becomes zero at $y=0$ and tends to the value $q_0$ at $y\rightarrow\pm\infty$. Thus $\vert q-q_0\vert$ has maximum equal to $q_0$ at $y=0$ and rapidly decreases outside domain wall}. 
    
    Let consider briefly the energetics of domain walls.  It follows from exact equation valid for bisector hDW's and is not associated with any variational ansatz:
    \begin{equation}
    E_{dw}=Jq_0\sqrt{I_1I_2}.
    \end{equation} 
    From previous analysis we find that at $\alpha\rightarrow 0$, the surface energy of the bisector hDW goes to zero as $\vert\alpha\vert^3$. At $\alpha=\rightarrow\pi$ this energy becomes infinite. This result seems to be inconsistent with equivalence of $\alpha=\pi$  and $\alpha=0$.  However, as we have shown earlier , just at the value of $\alpha=\pi$, $x-$component of wave vector is equal to zero at any $x$ and $y$, whereas its $y$-component turns into zero at the median of the hDW. Thus, the limit of the bisector hDW at $\alpha\rightarrow\pi$ leads to the specific minimization problem for a helical configuration with wave vector directed everywhere along $y-$axis, changing from $-q_0$ to $q_0$ and taking value 0 at $y=0$. Energy of such configuration is infinite as we have shown earlier.
    
    The micromagnetic calculations in \cite{Schoenherr 2018} qualitatively agree with our consideration in the range $0<\alpha<80$$^\circ$, though quantitatively this dependence is closer to $\alpha^2$  instead of our result $\alpha^3$. But in the second range of bisector hDW $140${$^\circ$}$<\alpha<180${$^\circ$}
    it results in monotonic decrease of surface energy to zero value at $\alpha=\pi$ in contrast to infinite surface energy predicted by our theory. The reason of these discrepancies presumably is the finite size of a "sample" used in micromagnetic calculations. It does not exceed $20\ell$.
    It becomes smaller than the width of the domain wall at $\alpha$ sufficiently close to 0 or $\pi$ invalidating the calculation of surface energy.  
    
    As for zig-zag domain walls we have seen
    already that at $\alpha=90$$^\circ$ and $\beta=0$, zig-zag line is symmetric,
    its sides have equal length. Therefore it could be expected that it
    realizes minimum of energy in comparison with either $\beta\neq0$
    or $\alpha\neq90$$^\circ$. Indeed such a minimum was found in the same micromagnetic
    calculations. However, energetically unfavorable configuration do
    not relax to this energy minimum. The metastability of these configurations
    may be associated with their complicated topological structure and
    small distances between disclinations and antidisclinations that increase
    their rigidity.
    
    In conclusion of this section we propose an approach to the problem of zig-zag domain walls at arbitrary angle $\alpha$
    between the wave vectors in the two domains based on its similarity with the devils
    staircase in the commenurate-incommensurate transition \cite{Pokrovsky 1983}. 
    Let consider a domain wall tilted with the angle $\gamma =\beta -\alpha/2$ to the direction of bisector.
    As we have seen already, such a tilt leads to a mismatch between periods of upper and
    lower helixes at their crossing with domain wall. These periods are $l_i=\ell /\cos(\alpha/2\mp\gamma);\,i=1,2$.
    The mismatch sums up to a full period if $(n+1)l_1=nl_2$ (we considered the simplest commensurate situation).
    This condition is satisfied if $\gamma=\gamma_n=\arctan[\cot(\frac{\alpha}{2n+1})]$. The mismatch can be avoided
    by introduction of a pair $\pm\pi$-disclinations. The distance between two such dipoles must be $L=n\ell$. 
    Let the distance between two disclination in the dipole is $d$. The
    mismatch though small ($q-q_0\propto\alpha\gamma q_0$ at small $\alpha$ and $\gamma$) persists at least at the distance $L$ generating
    the deviation  energy of the order $J(\gamma q_0\alpha d)^2/L$ per unit area. The rotation energy per unit area
    is $\sim J\alpha^2[\ln( L/d) ]/L$. Minimization over $d$ gives $d\sim\ell/\gamma^2$, but the energy decreases with $L$.
    We conclude that in the infinite system $L$ must be infinite at small angles $\alpha$ and $\beta$. Thus,
    we have proved that the domain wall tilted to bisector at small angle $\gamma$ are unstable.
    This stament agrees with experiment \cite{Schoenherr 2018} that did not observe such domain walls at small $\alpha$.
    The situation is different at large angles $\alpha$ and $\gamma$ since in this case the only scale of length for the 
    hDW is $L$. 
    
    { Sch\"onherr et al. \cite{Schoenherr 2018} argued theoretically and have shown numerically that surface energy of the disclination hDW has minimum at  angle between wave vectors $\alpha=\pi/2$ and $\beta=\gamma=0$. Our consideration shows that it should have more shallow minima at the same $\alpha$  and an infinite discrete set of $\gamma=\gamma_n$ corresponding to simple rational mismatch. Another rational mismatches of two steps of zig-zag and also zig-zags consisting of more than two steps would give a Devil's staircase of local minima, but only few of them with minimal denominators will be seen at finite temperature.}
    
    \section{Conclusions}
    Here we discuss open questions. Some of them are related to experiment. It would be very instructive to perform 
    measurements on the same or other samples, but at temperature lower than 260K in FeGe to ensure that the volume 
    anisotropy is more significant. Will arbitrary directions of wave vectors and angles between them persist?  
    It would be also interesting to apply polarized electron or neutron scattering to get information on the same 
    objects not only at the surface, but at least partly in the bulk. Though the samples used in experiment 
    \cite{Schoenherr 2018} were sufficiently large, there were no long regular hDW on presented pictures. This fact 
    interferes quantitative comparison of theory and experiment. 
    
    Among theoretical problems we would like to mention, first is the problem of the bulk-surface coupling or decoupling. 
    So far there is no satisfactory theory of this phenomenon. It is very important to develop variational methods for 
    structure and energy of the hDWs with the goal to get a desirable precision. So far we even could not compare our variational 
    calculations of bisector domain walls made for infinite sample with micromagnetic calculations in \cite{Schoenherr 2018} since the 
    latter were performed for finite and not too large samples. Their authors have found significant finite size effects. Thus, 
    it would be useful to develop variational approach to finite samples. Variational approach to theory of disclination domain 
    walls so far did not achieve quantiative level. We are looking now for a satisfactory trial function for magnetization.
    
    Important statements that were proved in this article include stability of the smooth (bisector) domain walls at all 
    angles between wave vectors and instability of zig-zag domain walls at small angles.

    \section{Acknowledgenents} This work was supported by the University of
    Cologne Center of Excellence QM2 and by William R. Thurman'58 Chair in Physics, Texas A\&M University. 
    We are thankful to Prof. D. Meier or very useful discussion of experimental procedure and sending us the original version of Fig. 5. Our thanks due to him and Prof. Y. Tokura for kind permission to use experimental figures from article \cite{Schoenherr 2018} 
    and to Dr. Chen Sun for his courteous help in preparation of Fig. 2.


\end{document}